\documentclass[prd,aps,floatfix,nofootinbib,11pt]{revtex4}
\usepackage{amsmath,epsfig}

\begin{document}

\title{$B\rightarrow 0^{+}\left( 1^{+}\right) +$ missing energy in
Unparticle Physics}
\author{M. Jamil Aslam$^{1,2}$}
\author{Cai-Dian L\"{u} $^{1}$}
\affiliation{$^{1}$Institute of High Energy Physics, P.O. Box 918(4), Beijing 100049,
China}
\affiliation{$^{2}$Department of Physics, COMSATS Institute of Information Technology,
Islamabad, Pakistan}

\begin{abstract}
We examine the effects of an unparticle $\mathcal{U}$ as a possible source
of missing energy in the $p$-wave decays of $B$ meson. The dependence of the
differential branching ratio on the $K_{0}^{\ast }\left( K_{1}\right) -$%
meson's energy is discussed in the presence of scalar and vector unparticle
operators and significant deviation from the standard model value is found
after addition of these operators. Finally, we have shown the dependence of
branching ratio for the above said decays on the parameters of the
unparticle stuff like the effective couplings, cutoff scale $\Lambda _{%
\mathcal{U}}$ and the scale dimensions $d_{\mathcal{U}}$.
\end{abstract}

\pacs{13.25.Hw; 12.90.+b} \maketitle





\section{  Introduction}

Flavor changing neutral current (FCNC) processes induced by $b\rightarrow s$
transitions are not allowed at tree level in the Standard Model (SM), but
are generated at loop level and are further suppressed by the CKM factors.
Therefore, these decays are very sensitive to the physics beyond the SM via
the influence of new particles in the loop. Though the branching ratios of
FCNC\ decays are small in the SM, quite interesting results are obtained
from the experiments both for the inclusive $B\rightarrow X_{s}\ell ^{+}\ell
^{-}$ \cite{1} and exclusive decay modes $B\rightarrow K\ell ^{+}\ell ^{-}$ %
\cite{2, 3, 4} and $B\rightarrow K^{\ast }\ell ^{+}\ell ^{-}$ \cite{5}.
These results are in good agreement with the theoretical estimates \cite{6,
7, 8}.

Among different semileptonic decays induced by \thinspace $b\rightarrow s$ \
transitions, $b\rightarrow s\nu \bar{\nu}$ decays are of particular
interest, because of the absence of a photonic penguin contribution and
hadronic long distance effects these have much smaller theoretical
uncertainties. But experimentally, it is too difficult to measure the
inclusive decay modes $B\rightarrow X_{s}\nu \bar{\nu}$ as one has to sum on
all the $X_{s}$'s. Therefore, the exclusive $B\rightarrow K\left( K^{\ast
}\right) \nu \bar{\nu}$ decays play a peculiar role both from the
experimental and theoretical point of view. The theoretical estimates of the
branching ratio of these decays are $Br\left( B\rightarrow K\nu \bar{\nu}%
\sim 10^{-5}\right) $ and $Br\left( B\rightarrow K^{\ast }\nu \bar{\nu}\sim
10^{-6}\right) $ \cite{9} whereas, the experimental bounds given by the $B$%
-factories, BELLE and BaBar, on these decays are \cite{10, 11}:
\begin{eqnarray}
Br\left( B\rightarrow K\nu \bar{\nu}\right) &<&1.4\times 10^{-5}
\label{exp-limits} \\
Br\left( B\rightarrow K^{\ast }\nu \bar{\nu}\right) &<&1.4\times 10^{-4}.
\notag
\end{eqnarray}%
These processes, based on $b\rightarrow s\nu \bar{\nu},$ are very sensitive
to the new physics and have been studied extensively in the literature in
the context of large extra dimension model and $Z^{\prime }$ models \cite%
{12, 13}. Any new physics model which can provide a relatively light new
source of missing energy (which is attributed to the neutrinos in the SM)
can potentially enhance the observed rates of $B\rightarrow K\left( K^{\ast
}\right) +$ missing energy. Recently, H. Georgi has proposed one such model
of Unparticles, which is one of the tantalizing issues these days \cite{14}.
The main idea of Georgi's model is that at a very high energy our theory
contains the fields of the standard model and the fields of a theory with a
nontrivial\ IR fixed point, which he called BZ (Banks-Zaks) fields \cite{15}%
. The interaction among the two sets is through the exchange of particles
with a large mass scale $M_{\mathcal{U}}$. The coupling between the SM
fields and BZ fields are nonrenormalizable below this scale and are
suppressed by the powers of $M_{\mathcal{U}}$. The renormalizable couplings
of the BZ fields then produce dimensional transmutation and the scale
invariant unparticle emerged below an energy scale $\Lambda _{\mathcal{U}}$.
In the effective theory below the scale $\Lambda _{\mathcal{U}}$ the BZ
operators matched onto unparticle operators, and the renormalizable
interaction matched onto a new set of interactions between standard model
and unparticle fields. The outcome of this model is the collection of
unparticle stuff with scale dimension $d_{\mathcal{U}}$, which is just like
a non-integral number of invisible massless particles, whose production
might be detectable in missing energy and momentum distributions \cite{16}.

This idea promoted a lot of interest in unparticle physics and its
signatures have been discussed at colliders \cite{16, 17, 17a, 18, 18a}, in
low energy physics \cite{19}, Lepton Flavor Violation \cite{20}, unparticle
physics effects in $B_{s}$ mixing \cite{20a}, and also in cosmology and
astrophysics \cite{21}. Aliev et al. have studied $B\rightarrow K\left(
K^{\ast }\right) +$ missing energy in unparticle physics \cite{22} in which
they have studied the effects of an unparticle $\mathcal{U}$ as a possible
source of a missing energy in these decays. They have found the dependence
of the differential branching ratio on the $K\left( K^{\ast }\right) $%
-meson's energy in the presence of scalar and vector unparticle operators
and then using the upper bounds on these decays, they put stringent
constraints on the parameters of the unparticle stuff.

The studies are even more complete if similar studies for the p-wave
decays of $B$ meson such as $B\rightarrow K_{0}^{\ast }\left(
1430\right) +\not\!\!{E}$
($\not\!\!{E}$ is missing energy) and $B\rightarrow K_{1}\left( 1270\right) +\not\!\!%
{E}$, where $K_{0}^{\ast }\left( 1430\right) $ \ and $K_{1}\left(
1270\right) $ are the pseudoscalar and axial vector mesons
respectively, carried out. In this paper, we have studied these
p-wave decays of $B$ meson in unparticle physics using the frame
work of Aliev et al. \cite{22}. We have considered the decay
$B\rightarrow K_{0}^{\ast }\left( K_{1}\right) \nu \bar{\nu}$ in SM
although for these modes no signals have been observed so for, but
in future B-factories where enough data is expected, these decays
will be observed. These SuperB factories will be measuring these
processes by analyzing the spectra of the final state hadron. In
doing this measurement a cut for high momentum on the hadron is
imposed, in order to suppress the background. Therefore, unparticle
would give us a unique distribution for the high energy hadron in
the final state, such that in future B-factories one will be able to
distinguish the presence of unparticle by observing the spectrum of
final state hadrons in $B\rightarrow \left( K,\ K^{\ast
},~K_{0}^{\ast },~K_{1}\right) +\not\!\!{E}$ \cite{22}.

The work is organized as follows: In section II after giving the expression
for the effective Hamiltonian for the decay $b\rightarrow s\nu \bar{\nu}$,
we define the scalar and vector unparticle physics operators for $%
b\rightarrow s\mathcal{U}$. Then using these expressions we
calculate the various contributions the decay rates of $B\rightarrow
K_{0}^{\ast }\left( K_{1}\right) +\not\!\!{E}$ both from the SM and
unparticle theory in Section III. Recently, Grinstein et al. gave
comments on the unparticle \cite{23} mentioning that Mack's
unitarity constraint lower bounds on CFT operator dimensions, e.g
$d_{\mathcal{U}}\geq 3$ for primary, gauge invariant, vector
unparticle operators. To account for this they have corrected the
results in the literature, and modified the propagator of vector and
tensor unparticles. We will also give the expressions of decay rate
using these modified vector operators in the same section. Finally,
section IV contains our numerical results and conclusions.

\section{Effective Hamiltonian in  SM and Unparticle operators}

The flavor changing neutral current $b\rightarrow s\nu \bar{\nu}$ are of
particular interest both from theoretical and experimental view. One of the
main reason of interest is the absence of long distance contribution related
to the four-quark operators in the effective Hamiltonian. In this respect,
the transition to neutrino represents a clean process even in comparison
with the $b\rightarrow s\gamma $ decay, where long-distance contributions,
though small, are expected to present \cite{24}. In Standard Model these
processes are governed by the effective Hamiltonian

\begin{equation}
\mathcal{H}_{eff}=\frac{G_{F}}{\sqrt{2}}\frac{\alpha }{2\pi }%
V_{tb}V_{ts}^{\ast }C_{10}\bar{s}\gamma ^{\mu }\left( 1-\gamma _{5}\right) b%
\bar{\nu}\gamma _{\mu }\left( 1-\gamma _{5}\right) \nu  \label{01}
\end{equation}%
where $V_{tb}V_{ts}^{\ast }$ are the elements of the
Cabbibo-Kobayashi Maskawa Matrix and $C_{10}$ is obtained from the
$Z^{0}$ penguin and box diagrams where the dominant contribution
corresponds to a top quark intermediate state and it is
\begin{equation}
C_{10}=\frac{D\left( x_{t}\right) }{\sin ^{2}\theta _{w}}.  \label{02}
\end{equation}%
$\theta _{w}$ is the Weinberg angle and $D\left( x_{t}\right) $ is the usual
Inami-Lim function, given as
\begin{equation}
D\left( x_{t}\right) =\frac{x_{t}}{8}\left\{ \frac{x_{t}+2}{x_{t}-1}+\frac{%
3x_{t}-6}{\left( x_{t}-1\right) ^{2}}\ln \left( x_{t}\right) \right\} ,
\label{03}
\end{equation}%
with $x_{t}=m_{t}^{2}/m_{W}^{2}$.

The unparticle transition at the quark level can be described by $%
b\rightarrow s\mathcal{U}$, where one can consider the following operators:

\begin{itemize}
\item Scalar unparticle operator
\begin{equation}
C_{s}\frac{1}{\Lambda _{\mathcal{U}}^{d_{\mathcal{U}}}}\bar{s}\gamma _{\mu
}b\partial ^{\mu }O_{\mathcal{U}}+C_{P}\frac{1}{\Lambda _{\mathcal{U}}^{d_{%
\mathcal{U}}}}\bar{s}\gamma _{\mu }\gamma _{5}b\partial ^{\mu }O_{\mathcal{U}%
}  \label{04}
\end{equation}

\item Vector unparticle operator
\begin{equation}
C_{V}\frac{1}{\Lambda _{\mathcal{U}}^{d_{\mathcal{U}}}}\bar{s}\gamma _{\mu
}bO_{\mathcal{U}}^{\mu }+C_{A}\frac{1}{\Lambda _{\mathcal{U}}^{d_{\mathcal{U}%
}}}\bar{s}\gamma _{\mu }\gamma _{5}bO_{\mathcal{U}}^{\mu }.  \label{05}
\end{equation}
\end{itemize}

The propagator for the scalar unparticle field can be written as\cite{14,
16, 17}
\begin{equation}
\int d^{4}xe^{iP\cdot x}\left\langle 0\left| TO_{\mathcal{U}}\left( x\right)
O_{\mathcal{U}}\left( 0\right) \right| 0\right\rangle =i\frac{A_{d_{_{%
\mathcal{U}}}}}{2\sin \left( d_{_{\mathcal{U}}}\pi \right) }\left(
-P^{2}\right) ^{d_{_{\mathcal{U}}}-2}  \label{06}
\end{equation}%
with
\begin{equation}
A_{d_{_{\mathcal{U}}}}=\frac{16\pi ^{5/2}}{\left( 2\pi \right) ^{2d_{_{%
\mathcal{U}}}}}\frac{\Gamma \left( d_{_{\mathcal{U}}}+1/2\right) }{\Gamma
\left( d_{_{\mathcal{U}}}-1\right) \Gamma \left( 2d_{_{\mathcal{U}}}\right) }%
.  \label{07}
\end{equation}

\section{Differential Decay Widths}

In Standard Model the decay $B\rightarrow K_{0}^{\ast }\left( K_{1}\right) +%
\not\!\!{E}$ is described by the decay $B\rightarrow K_{0}^{\ast
}\left( K_{1}\right) \nu \bar{\nu}$. At quark level this process is
governed by the effective Hamiltonian defined in Eq. (\ref{01})
which when sandwiched between $B$ and $K_{0}^{\ast }\left(
K_{1}\right) $ involves the hadronic matrix elements for the
exclusive decay $B\rightarrow K_{0}^{\ast }\left( K_{1}\right) \nu
\bar{\nu}$. These can be parameterized by the form factors
and the non-vanishing matrix elements for $B\rightarrow K_{0}^{\ast }$ are %
\cite{24}
\begin{equation}
\left\langle K_{0}^{\ast }\left( p^{\prime }\right) \left| \bar{s}\gamma
_{\mu }\gamma _{5}b\right| B\left( p\right) \right\rangle =-i\left[
f_{+}\left( q^{2}\right) \left( p+p^{\prime }\right) _{\mu }+f_{-}\left(
q^{2}\right) q_{\mu }\right] .  \label{08}
\end{equation}%
where $q_{\mu }=\left( p+p^{\prime }\right) _{\mu }$. Using the above
definition and taking into account the three species of neutrinos in the
Standard Model, the differential decay width as a function of $K_{0}^{\ast }$
energy $\left( E_{K_{0}^{\ast }}\right) $ can be written as \cite{24}:
\begin{equation}
\frac{d\Gamma ^{SM}}{dE_{K_{0}^{\ast }}}=\frac{G_{F}^{2}\alpha ^{2}}{%
2^{7}\pi ^{5}M_{B}^{2}}\left| V_{tb}V_{ts}^{\ast }\right| ^{2}\left|
C_{10}\right| ^{2}f_{+}^{2}\left( q^{2}\right) \sqrt{\lambda ^{3}\left(
M_{B}^{2},M_{K_{0}^{\ast }}^{2},q^{2}\right) }  \label{09}
\end{equation}%
with $\lambda \left( a,b,c\right) =a^{2}+b^{2}+c^{2}-2ab-2bc-2ca$ and $%
q^{2}=M_{B}^{2}+M_{K_{0}^{\ast }}^{2}-2M_{B}E_{K_{0}^{\ast }}$ . Here $%
f_{+}\left( q^{2}\right) $ and $f_{-}\left( q^{2}\right) $ are the form
factors which are the non-perturbative quantities and can be calculated
using some models. The one we have used here was calculated by using Light
Front Quark Model (LFQR) by Cheng et al. \cite{24} and these can be
parameterized as:%
\begin{equation*}
F\left( q^{2}\right) =\frac{F\left( 0\right) }{1-aq^{2}/M_{B}^{2}+b\left(
q^{2}/M_{B}^{2}\right) ^{2}}
\end{equation*}%
and the fitted parameters are given in Table~\ref{di-fit}.

\begin{table}[tbh]
\caption{{}The parameters for $B\rightarrow K_{0}^{\ast }$ form factors.}
\label{di-fit}%
\begin{tabular}{cccc}
\hline\hline
& $F\left( 0\right) $ & $a$ & $b$ \\ \hline
$f_{+}$ & $-0.26$ & $1.36$ & $0.86$ \\
$f_{-}$ & $0.21$ & $1.26$ & $0.93$ \\ \hline\hline
\end{tabular}%
\end{table}

Similarly, for $B\rightarrow K_{1}$ transition the matrix elements can be
parametrized as \cite{25}
\begin{eqnarray}
\left\langle K_{1}(k,\varepsilon )\left| V_{\mu }\right| B(p)\right\rangle
&=&i\varepsilon _{\mu }^{\ast }\left( M_{B}+M_{K_{1}}\right) V_{1}(q^{2})
\notag \\
&&-(p+k)_{\mu }\left( \varepsilon ^{\ast }\cdot q\right) \frac{V_{2}(q^{2})}{%
M_{B}+M_{K_{1}}}  \notag \\
&&-q_{\mu }\left( \varepsilon \cdot q\right) \frac{2M_{K_{1}}}{s}\left[
V_{3}(q^{2})-V_{0}(q^{2})\right]  \label{10} \\
\left\langle K_{1}(k,\varepsilon )\left| A_{\mu }\right| B(p)\right\rangle
&=&\frac{2i\epsilon _{\mu \nu \alpha \beta }}{M_{B}+M_{K_{1}}}\varepsilon
^{\ast \nu }p^{\alpha }k^{\beta }A(q^{2})  \label{11}
\end{eqnarray}%
where $V_{\mu }=\bar{s}\gamma _{\mu }b$ and $A_{\mu }=\bar{s}\gamma _{\mu
}\gamma _{5}b$ are the vector and axial vector currents respectively and $%
\varepsilon _{\mu }^{\ast }$ is the polarization vector for the final state
axial vector meson. In this case we have used the form factors that were
calculated by Paracha et al. \cite{25} and the corresponding expressions are:%
\begin{eqnarray}
A\left( s\right) &=&\frac{A\left( 0\right) }{\left( 1-s/M_{B}^{2}\right)
(1-s/M_{B}^{\prime 2})}  \notag \\
V_{1}(s) &=&\frac{V_{1}(0)}{\left( 1-s/M_{B_{A}^{\ast }}^{2}\right) \left(
1-s/M_{B_{A}^{\ast }}^{\prime 2}\right) }\left( 1-\frac{s}{%
M_{B}^{2}-M_{K_{1}}^{2}}\right)  \label{form-factors} \\
V_{2}(s) &=&\frac{\tilde{V}_{2}(0)}{\left( 1-s/M_{B_{A}^{\ast }}^{2}\right)
\left( 1-s/M_{B_{A}^{\ast }}^{\prime 2}\right) }-\frac{2M_{K_{1}}}{%
M_{B}-M_{K_{1}}}\frac{V_{0}(0)}{\left( 1-s/M_{B}^{2}\right) \left(
1-s/M_{B}^{\prime 2}\right) }  \notag
\end{eqnarray}%
with
\begin{eqnarray}
A(0) &=&-(0.52\pm 0.05)  \notag \\
V_{1}(0) &=&-(0.24\pm 0.02)  \notag \\
\tilde{V}_{2}(0) &=&-(0.39\pm 0.03).  \label{Num-f-factor}
\end{eqnarray}%
The differential decay rate can be calculated as \cite{22}:
\begin{equation}
\frac{d\Gamma ^{SM}}{dE_{K_{1}}}=\frac{G_{F}^{2}\alpha ^{2}}{2^{9}\pi
^{5}M_{B}^{2}}\left| V_{tb}V_{ts}^{\ast }\right| ^{2}\left| C_{10}\right|
^{2}\lambda ^{1/2}\left| M_{SM}\right| ^{2}  \label{12}
\end{equation}%
where%
\begin{eqnarray}
\left| M_{SM}\right| ^{2} &=&\frac{8q^{2}\lambda \left| A\left( q^{2}\right)
\right| ^{2}}{\left( M_{B}+M_{K_{1}}\right) ^{2}}+\frac{1}{M_{K_{1}}^{2}}%
\bigg[\lambda ^{2}\frac{\left| V_{2}\left( q^{2}\right) \right| ^{2}}{\left(
M_{B}+M_{K_{1}}\right) ^{2}}+\left( M_{B}+M_{K_{1}}\right) ^{2}\left(
\lambda +12M_{K_{1}}^{2}q^{2}\right) \left| V_{1}\left( q^{2}\right) \right|
^{2}  \notag \\
&&-\lambda \left( M_{B}^{2}-M_{K_{1}}^{2}-q^{2}\right)
\rm{Re}(V_{1}^{\ast }\left( q^{2}\right) V_{2}\left( q^{2}\right)
+V_{2}^{\ast }\left( q^{2}\right) V_{1}\left( q^{2}\right) )\bigg]
\label{13a}
\end{eqnarray}
and $\lambda =\lambda \left( M_{B}^{2},M_{K_{1}}^{2},q^{2}\right) $ with $%
q^{2}=M_{B}^{2}+M_{K_{1}}^{2}-2M_{B}E_{K_{1}}$.

Now in decay mode $B\rightarrow K_{0}^{\ast }\left( K_{1}\right)
+\not\!\!{E}$, the missing energy shown by $\not\!\!{E}$ can also be
attributed to the unparticle and hence the unparticle can also
contribute to these decay modes. Therefore, the signature of two
decay modes $B\rightarrow K_{0}^{\ast }\left( K_{1}\right) \nu
\bar{\nu}$ and $B\rightarrow K_{0}^{\ast }\left( K_{1}\right)
\mathcal{U}$ is required like the one done for $B\rightarrow K\left(
K^{\ast }\right) \nu \bar{\nu}$ and $B\rightarrow K\left( K^{\ast
}\right) \mathcal{U}$ in the literature \cite{22}.

\subsection{The Scalar Unparticle Operator}

Using the scalar unparticle operator defined in Eq. (\ref{04}) the matrix
element for $B\rightarrow K_{0}^{\ast }\mathcal{U}$ can be written as
\begin{eqnarray}
\mathcal{M}_{K_{0}^{\ast }}^{S\mathcal{U}} &=&\frac{1}{\Lambda ^{d_{\mathcal{%
U}}}}\left\langle K_{0}^{\ast }\left( p^{\prime }\right) \left| \bar{s}%
\gamma _{\mu }\left( \mathcal{C}_{S}+\mathcal{C}_{P}\gamma _{5}\right)
b\right| B\left( p\right) \right\rangle \partial ^{\mu }O_{\mathcal{U}}
\notag \\
&=&\frac{1}{\Lambda ^{d_{\mathcal{U}}}}\mathcal{C}_{P}[f_{+}\left(
q^{2}\right) \left( M_{B}^{2}-M_{K_{0}^{\ast }}^{2}\right) +f_{-}\left(
q^{2}\right) q^{2}]O_{\mathcal{U}}  \label{14}
\end{eqnarray}%
Now the decay rate for $B\rightarrow K_{0}^{\ast }\mathcal{U}$ can be
evaluated to be:
\begin{equation}
\frac{d\Gamma ^{S\mathcal{U}}}{dE_{K_{0}^{\ast }}}=\frac{1}{8\pi ^{2}m_{B}}%
\sqrt{E_{K_{0}^{\ast }}^{2}-M_{K_{0}^{\ast }}^{2}}\left| \mathcal{M}^{S%
\mathcal{U}}\right| ^{2}  \label{14a}
\end{equation}%
where
\begin{eqnarray}
\left| \mathcal{M}^{S\mathcal{U}}\right| ^{2} &=&\left| \mathcal{C}%
_{P}\right| ^{2}\frac{A_{d_{\mathcal{U}}}}{\Lambda ^{^{2d_{\mathcal{U}}}}}%
\left( M_{B}^{2}+M_{K_{0}^{\ast }}^{2}-2M_{B}E_{K_{0}^{\ast }}\right) ^{d_{%
\mathcal{U}}-2}  \label{15} \\
&&\left. \times \left[ f_{+}\left( q^{2}\right) \left(
M_{B}^{2}-M_{K_{0}^{\ast }}^{2}\right) +f_{-}\left( q^{2}\right) \left(
M_{B}^{2}+M_{K_{0}^{\ast }}^{2}-2M_{B}E_{K_{0}^{\ast }}\right) \right]
^{2}\right. .  \notag
\end{eqnarray}%
Following the same lines, the corresponding matrix element $B\rightarrow
K_{1}\mathcal{U}$ is
\begin{eqnarray}
\mathcal{M}_{K_{1}}^{S\mathcal{U}} &=&\frac{1}{\Lambda ^{d_{\mathcal{U}}}}%
\left\langle K_{1}\left( p^{\prime }\right) \left| \bar{s}\gamma _{\mu
}\left( \mathcal{C}_{S}+\mathcal{C}_{P}\gamma _{5}\right) b\right| B\left(
p\right) \right\rangle \partial ^{\mu }O_{\mathcal{U}}  \notag \\
&=&\frac{i}{\Lambda ^{d_{\mathcal{U}}}}\mathcal{C}_{S}\left( \varepsilon
^{\ast }\cdot q\right) \bigg[(M_{B}+M_{K_{1}})V_{1}\left( q^{2}\right)
\notag \\
&&-(M_{B}-M_{K_{1}})V_{2}\left( q^{2}\right) -2M_{K_{1}}\left( V_{3}\left(
q^{2}\right) -V_{0}\left( q^{2}\right) \right) \bigg]O_{\mathcal{U}},
\label{16a}
\end{eqnarray}%
and the differential decay rate is
\begin{equation}
\frac{d\Gamma ^{S\mathcal{U}}}{dE_{K_{1}}}=\frac{M_{B}}{2\pi ^{2}}\frac{%
A_{d_{\mathcal{U}}}}{\Lambda ^{^{2d_{\mathcal{U}}}}}\left| \mathcal{C}%
_{S}\right| ^{2}\left| V_{0}\left( q^{2}\right) \right| ^{2}\left(
E_{K_{1}}^{2}-M_{K_{1}}^{2}\right) ^{3/2}\left(
M_{B}^{2}+M_{K_{1}}^{2}-2M_{B}E_{K_{1}}\right) ^{d_{\mathcal{U}}-2}.
\label{17}
\end{equation}

One can see from Eq. (\ref{14a}) and Eq. (\ref{17}) that the scalar
unparticle contribution to the decay rate depends on $\mathcal{C}_{P}$, $%
\mathcal{C}_{S}$, $d_{\mathcal{U}}$ and $\Lambda _{\mathcal{U}}$, therefore
one can see the behavior of decay rates for the said decays on these
parameters which will be hoped to get constraint once we have experimental
data on these decays. This we will do in a separate section.

\subsection{The Vector Unparticle Operator}

The matrix element for $B\rightarrow K_{0}^{\ast }\mathcal{U}$ using the
vector unparticle operator defined in Eq. (\ref{05}) and the definition of
form factors given in Eq. (\ref{08}) can be calculated as:
\begin{eqnarray}
\mathcal{M}_{K_{0}^{\ast }}^{V\mathcal{U}} &=&\frac{1}{\Lambda ^{d_{\mathcal{%
U}}-1}}\left\langle K_{0}^{\ast }\left( p^{\prime }\right) \left| \bar{s}%
\gamma _{\mu }\left( \mathcal{C}_{V}+\mathcal{C}_{A}\gamma _{5}\right)
b\right| B\left( p\right) \right\rangle O_{\mathcal{U}}^{\mu }  \notag \\
&=&\frac{1}{\Lambda ^{d_{\mathcal{U}}-1}}\mathcal{C}_{A}[f_{+}\left(
q^{2}\right) \left( p+p^{\prime }\right) _{\mu }+f_{-}\left( q^{2}\right)
q_{\mu }]O_{\mathcal{U}}^{\mu }.  \label{18}
\end{eqnarray}%
The differential decay rate is then
\begin{eqnarray}
\frac{d\Gamma ^{V\mathcal{U}}}{dE_{K_{0}^{\ast }}} &=&\frac{1}{8\pi ^{2}m_{B}%
}\frac{A_{d_{\mathcal{U}}}}{\Lambda ^{^{2d_{\mathcal{U}}-2}}}\left| \mathcal{%
C}_{A}\right| ^{2}\left| f_{+}\left( q^{2}\right) \right| ^{2}\left(
M_{B}^{2}+M_{K_{0}^{\ast }}^{2}-2M_{B}E_{K_{0}^{\ast }}\right) ^{d_{\mathcal{%
U}}-2}\sqrt{E_{K_{0}^{\ast }}^{2}-M_{K_{0}^{\ast }}^{2}}  \notag \\
&&\times \left\{ -\left( M_{B}^{2}+M_{K_{0}^{\ast
}}^{2}+2M_{B}E_{K_{0}^{\ast }}\right) +\frac{\left( M_{B}^{2}-M_{K_{0}^{\ast
}}^{2}\right) ^{2}}{\left( M_{B}^{2}+M_{K_{0}^{\ast
}}^{2}-2M_{B}E_{K_{0}^{\ast }}\right) }\right\} .  \label{19}
\end{eqnarray}%
For $B\rightarrow K_{1}$ case the matrix element for $B\rightarrow K_{1}%
\mathcal{U~}$is
\begin{eqnarray}
\mathcal{M}_{K_{1}}^{V\mathcal{U}} &=&\frac{1}{\Lambda ^{d_{\mathcal{U}}-1}}%
\left\langle K_{1}\left( p^{\prime }\right) \left| \bar{s}\gamma _{\mu
}\left( \mathcal{C}_{V}+\mathcal{C}_{A}\gamma _{5}\right) b\right| B\left(
p\right) \right\rangle O_{\mathcal{U}}^{\mu }  \notag \\
&=&\bigg[\frac{\mathcal{C}_{V}}{\Lambda ^{d_{\mathcal{U}}-1}}(i\varepsilon
_{\mu }^{\ast }\left( M_{B}+M_{K_{1}}\right) V_{1}\left( q^{2}\right)
-i\left( p+p^{\prime }\right) _{\mu }\left( \varepsilon ^{\ast }\cdot
q\right) \frac{V_{2}\left( q^{2}\right) }{M_{B}+M_{K_{1}}}  \notag \\
&&  \label{19a} \\
&&-iq_{\mu }\left( \varepsilon ^{\ast }\cdot q\right) \frac{2M_{K_{1}}}{q^{2}%
}\left( V_{3}\left( q^{2}\right) -V_{0}\left( q^{2}\right) \right) )+\frac{%
\mathcal{C}_{A}}{\Lambda ^{d_{\mathcal{U}}-1}}(\frac{2A\left( q^{2}\right) }{%
M_{B}+M_{K_{1}}}\epsilon _{\mu \nu \alpha \beta }\varepsilon ^{\nu \ast
}p^{\alpha }p^{\prime \beta })\bigg]O_{\mathcal{U}}^{\mu }  \notag
\end{eqnarray}%
and the differential decay rate will be:
\begin{eqnarray}
\frac{d\Gamma ^{V\mathcal{U}}}{dE_{K_{1}}} &=&\frac{1}{8\pi ^{2}m_{B}}\frac{%
A_{d_{\mathcal{U}}}}{\Lambda ^{^{2d_{\mathcal{U}}-2}}}\sqrt{%
E_{K_{1}}^{2}-M_{K_{1}}^{2}}\left( q^{2}\right) ^{d_{\mathcal{U}}-2}  \notag
\\
&&\bigg[8\left| \mathcal{C}_{A}\right| ^{2}M_{B}^{2}\left(
E_{K_{1}}^{2}-M_{K_{1}}^{2}\right) \frac{A\left( q^{2}\right) }{\left(
M_{B}+M_{K_{1}}\right) ^{2}}  \notag \\
&&+\left| \mathcal{C}_{V}\right| ^{2}\frac{1}{M_{K_{1}}^{2}\left(
M_{B}+M_{K_{1}}\right) ^{2}q^{2}}  \notag \\
&&\bigg[\left( M_{B}+M_{K_{1}}\right) ^{4}\left(
3M_{K_{1}}^{4}+2M_{B}^{2}M_{K_{1}}^{2}-6M_{B}M_{K_{1}}^{2}E_{K_{1}}+M_{B}^{2}E_{K_{1}}^{2}\right) \left| V_{1}\left( q^{2}\right) \right| ^{2}
\notag \\
&&+2M_{B}^{4}\left( E_{K_{1}}^{2}-M_{K_{1}}^{2}\right) \left| V_{2}\left(
q^{2}\right) \right| ^{2}+4\left( M_{B}+M_{K_{1}}\right) ^{2}  \notag \\
&&\left( M_{B}E_{K_{1}}-M_{K_{1}}^{2}\right) \left(
M_{K_{1}}^{2}-E_{K_{1}}^{2}\right) M_{B}^{2}\left( V_{1}V_{2}^{\ast
}+V_{2}V_{1}^{\ast }\right) \bigg]\bigg]  \label{19b}
\end{eqnarray}%
The total decay width can be obtained if we integrate on the energy
of the final state meson in the range $M_{K\left( K_{1}\right)
}<E_{K\left( K_{1}\right) }<\left( M_{B}^{2}+M_{K\left( K_{1}\right)
}^{2}\right) /2M_{B}$ for $B\rightarrow K\left( K_{1}\right)
+\not\!\!{E}$.

Recently, Grinstein et al. have given comment on the unparticle \cite{23} in
which they have mentioned that Mack's unitarity constraint lower bounds \ on
CFT operator dimensions, e.g. $d_{\mathcal{U}}\geq 3$ for primary, gauge
invariant, vector unparticle operators. To account for this they have
corrected the results in the literature, and modified the propagator of
vector and tensor unparticles. The modified vector propagator is
\begin{equation}
\int d^{4}xe^{iPx}\left\langle 0\left| T\left( O_{\mathcal{U}}^{\mu }\left(
x\right) O_{\mathcal{U}}^{\nu }\left( x\right) \right) \right|
0\right\rangle =A_{d_{\mathcal{U}}}\left( -g^{\mu \nu }+aP^{\mu }P^{\nu
}/P^{2}\right) \left( P^{2}\right) ^{d_{\mathcal{U}}-2}.  \label{21}
\end{equation}%
Here $P$ is the momentum of the unparticle, $A_{d_{\mathcal{U}}}$ is defined
in Eq. (\ref{07}) \ and $a\neq 1($ in contrast to the value $a=1$ which was
considered by Georgi \cite{14}) but is defined as:
\begin{equation}
a=\frac{2\left( d_{\mathcal{U}}-2\right) }{\left( d_{\mathcal{U}}-1\right) }.
\label{22}
\end{equation}%
By incorporating this factor $a$ in the vector unparticle operator the Eqs. (%
\ref{19}) and (\ref{19b}) get modification and the modified result of the
decay rate for $B\rightarrow K_{0}^{\ast }\mathcal{U}$ is
\begin{eqnarray}
\frac{d\Gamma ^{V\mathcal{U}}}{dE_{K_{0}^{\ast }}} &=&\frac{1}{8\pi ^{2}m_{B}%
}\frac{A_{d_{{}}\mathcal{U}}}{\Lambda ^{^{2d_{\mathcal{U}}-2}}}\left|
\mathcal{C}_{A}\right| ^{2}\left| f_{+}\left( q^{2}\right) \right|
^{2}\left( M_{B}^{2}+M_{K_{0}^{\ast }}^{2}-2M_{B}E_{K_{0}^{\ast }}\right)
^{d_{\mathcal{U}}-2}\sqrt{E_{K_{0}^{\ast }}^{2}-M_{K_{0}^{\ast }}^{2}}
\notag \\
&&\bigg[\left| f_{+}\left( q^{2}\right) \right| ^{2}\left( -\left(
M_{B}^{2}+M_{K_{0}^{\ast }}^{2}+2M_{B}E_{K_{0}^{\ast }}\right) +\frac{%
a\left( M_{B}^{2}-M_{K_{0}^{\ast }}^{2}\right) ^{2}}{\left(
M_{B}^{2}+M_{K_{0}^{\ast }}^{2}-2M_{B}E_{K_{0}^{\ast }}\right) }\right)
\notag \\
&&+\left| f_{-}\left( q^{2}\right) \right| ^{2}\left( a-1\right) \left(
M_{B}^{2}+M_{K_{0}^{\ast }}^{2}-2M_{B}E_{K_{0}^{\ast }}\right)  \notag \\
&&+2\left( a-1\right) \left( f_{+}\left( q^{2}\right) f_{-}\left(
q^{2}\right) \right) \left( M_{B}^{2}-M_{K_{0}^{\ast }}^{2}\right) \bigg]
\notag \\
&&  \label{23}
\end{eqnarray}%
Similarly, for $B\rightarrow K_{1}\mathcal{U}$ the result becomes
\begin{eqnarray}
\frac{d\Gamma ^{V\mathcal{U}}}{dE_{K_{1}}} &=&\frac{1}{8\pi ^{2}m_{B}}\frac{%
A_{d_{\mathcal{U}}}}{\Lambda ^{^{2d_{\mathcal{U}}-2}}}\sqrt{%
E_{K_{1}}^{2}-M_{K_{1}}^{2}}\left( q^{2}\right) ^{d_{\mathcal{U}}-2}  \notag
\\
&&\bigg[\left| \mathcal{M}_{11}\right| ^{2}+\left| \mathcal{M}_{22}\right|
^{2}+\left| \mathcal{M}_{33}\right| ^{2}+\left| \mathcal{M}_{44}\right|
^{2}+\left| \mathcal{M}_{23}\right| ^{2}+\left| \mathcal{M}_{24}\right|
^{2}+\left| \mathcal{M}_{34}\right| ^{2}\bigg]  \notag \\
&&  \label{24}
\end{eqnarray}%
with%
\begin{eqnarray*}
\left| \mathcal{M}_{11}\right| ^{2} &=&8\left| \mathcal{C}_{A}\right|
^{2}M_{B}^{2}\left( E_{K_{1}}^{2}-M_{K_{1}}^{2}\right) \frac{A\left(
q^{2}\right) }{\left( M_{B}+M_{K_{1}}\right) ^{2}} \\
\left| \mathcal{M}_{22}\right| ^{2} &=&\left| \mathcal{C}_{V}\right| ^{2}%
\frac{1}{M_{K_{1}}^{2}\left( M_{B}+M_{K_{1}}\right) ^{2}q^{2}} \\
&&\bigg[\left( M_{B}+M_{K_{1}}\right) ^{4}\left( 3M_{K_{1}}^{2}\left(
M_{B}^{2}+M_{K_{1}}^{2}-2M_{B}E_{K_{1}}\right) -a\left(
M_{B}^{2}M_{K_{1}}^{2}-M_{B}^{2}E_{K_{1}}^{2}\right) \right) \left|
V_{1}\left( q^{2}\right) \right| ^{2}\bigg] \\
\left| \mathcal{M}_{33}\right| ^{2} &=&\left| \mathcal{C}_{V}\right| ^{2}%
\frac{1}{M_{K_{1}}^{2}\left( M_{B}+M_{K_{1}}\right) ^{2}q^{2}} \\
&&\bigg[M_{B}^{2}\left( E_{K_{1}}^{2}-M_{K_{1}}^{2}\right) \left( a\left(
M_{B}^{2}-M_{K_{1}}^{2}\right) ^{2}+\left( 2M_{B}E_{K_{1}}\right)
^{2}-\left( M_{B}^{2}+M_{K_{1}}^{2}\right) ^{2}\right) \left| V_{2}\left(
q^{2}\right) \right| ^{2}\bigg] \\
\left| \mathcal{M}_{44}\right| ^{2} &=&\left| \mathcal{C}_{V}\right| ^{2}%
\frac{1}{M_{K_{1}}^{2}\left( M_{B}+M_{K_{1}}\right) ^{2}q^{2}} \\
&&\bigg[4M_{B}^{2}\left( M_{B}+M_{K_{1}}\right) ^{2}\left(
E_{K_{1}}^{2}-M_{K_{1}}^{2}\right) \left( a-1\right) M_{K_{1}}^{2}\left|
V_{3}\left( q^{2}\right) -V_{0}\left( q^{2}\right) \right| ^{2}\bigg]
\end{eqnarray*}%
\begin{eqnarray}
\left| \mathcal{M}_{23}\right| ^{2} &=&\left| \mathcal{C}_{V}\right| ^{2}%
\frac{1}{M_{K_{1}}^{2}\left( M_{B}+M_{K_{1}}\right) ^{2}q^{2}}  \notag \\
&&\bigg[M_{B}^{2}\left( M_{B}+M_{K_{1}}\right) ^{2}\left(
E_{K_{1}}^{2}-M_{K_{1}}^{2}\right) \left(
M_{B}^{2}+M_{K_{1}}^{2}-2M_{B}E_{K_{1}}-a\left(
M_{B}^{2}-M_{K_{1}}^{2}\right) \right)  \notag \\
&&\left( V_{1}\left( q^{2}\right) V_{2}^{\ast }\left( q^{2}\right)
+V_{2}\left( q^{2}\right) V_{1}^{\ast }\left( q^{2}\right) \right) \bigg]
\notag \\
\left| \mathcal{M}_{24}\right| ^{2} &=&\left| \mathcal{C}_{V}\right| ^{2}%
\frac{1}{M_{K_{1}}^{2}\left( M_{B}+M_{K_{1}}\right) ^{2}q^{2}}  \notag \\
&&\bigg[2M_{K_{1}}\left( M_{B}+M_{K_{1}}\right) ^{3}\left( \left( 1-a\right)
M_{B}^{2}\left( E_{K_{1}}^{2}-M_{K_{1}}^{2}\right) \right) \left(
V_{1}\left( V_{3}-V_{0}\right) ^{\ast }+\left( V_{3}-V_{0}\right)
V_{1}^{\ast }\right) \bigg]  \notag \\
\left| \mathcal{M}_{34}\right| ^{2} &=&\left| \mathcal{C}_{V}\right| ^{2}%
\frac{1}{M_{K_{1}}^{2}\left( M_{B}+M_{K_{1}}\right) ^{2}q^{2}}\bigg[%
2M_{K_{1}}\left( M_{B}+M_{K_{1}}\right) \left( M_{B}^{2}-M_{K_{1}}^{2}\right)
\notag \\
&&M_{B}^{2}\left( E_{K_{1}}^{2}-M_{K_{1}}^{2}\right) \left( a-1\right)
\left( V_{2}\left( V_{3}-V_{0}\right) ^{\ast }+\left( V_{3}-V_{0}\right)
V_{2}^{\ast }\right) \bigg]  \label{24amplitude}
\end{eqnarray}%
One can easily see that Eqs. (\ref{23}) and (\ref{24}) reduces to the Eqs. (%
\ref{19}) and (\ref{19b}) respectively, if one puts $a=1$.

\section{Results and Discussions}

In this section we present our numerical study for the $B\rightarrow
K_{0}^{\ast }\left( K_{1}\right) +\not\!\!{E}$ where we try to
distinguish the unparticle physics effects from that of the SM. In
Standard Model $\not\!\!{E}$ which is the missing energy is
attributed to the neutrinos where as in the case under
consideration, this is attached to the unparticle. Therefore the
total decay rate can be written as
\begin{equation}
\Gamma =\Gamma ^{SM}+\Gamma ^{\mathcal{U}}.  \label{25}
\end{equation}%
Here $\Gamma ^{SM}$ is the Standard Model contribution $\left( B\rightarrow
K_{0}^{\ast }\left( K_{1}\right) \nu \bar{\nu}\right) $ where as $\Gamma ^{%
\mathcal{U}}$ is from the unparticle $\left( B\rightarrow
K_{0}^{\ast }\left( K_{1}\right) \mathcal{U}\right) $ to the decay
$B\rightarrow K_{0}^{\ast }\left( K_{1}\right) +\not\!\!{E}.$ In
ref. \cite{22} it is pointed out that the SM\ process $B\rightarrow
K\left( K^{\ast }\right) \nu \bar{\nu} $ provides a unique energy
distribution spectrum of final state hadrons and present
experimental limits on the branching ratio of these processes are
about an order of magnitude below the respective SM expectation
values. They have used experimental upper limit on the branching
ratio of $B\rightarrow K\left( K^{\ast }\right) \nu \bar{\nu}$ decay
to estimate the constraints on the unparticle properties.

\begin{figure}[htb]
\begin{center}
\includegraphics[scale=0.7]{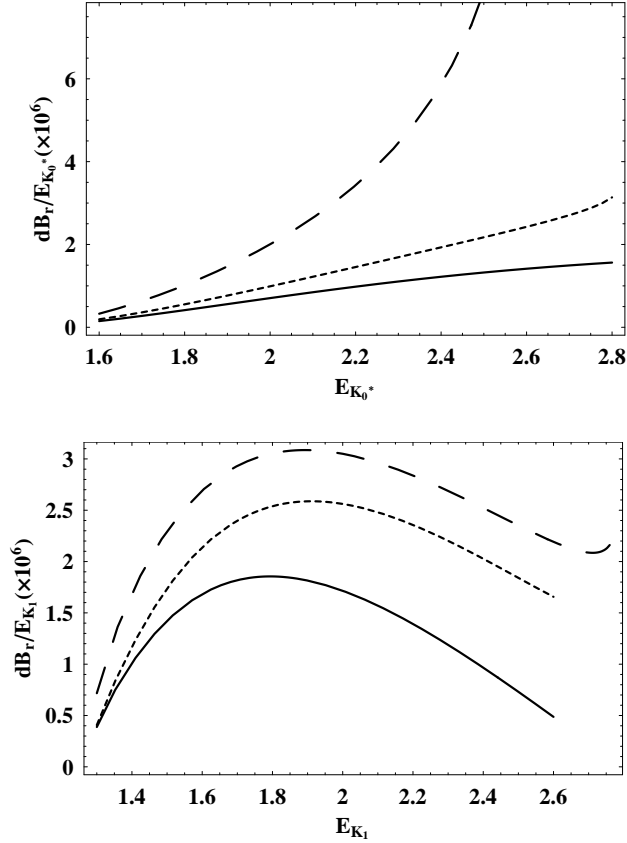}
\caption{The differential branching ratio for $B\rightarrow
K_{0}^{\ast }\left( K_{1}\right) +\not\!\!{E}$ as a function of
hadronic energy $E_{K_{0}^{\ast }}\left( E_{K_{1}}\right) $ is
plotted. Top panel is for $B\rightarrow K_{0}^{\ast }+\not\!\!{E}$
and bottom is for $B\rightarrow K_{1}+\not\!\!{E}$. The other
parameters are $d_{\mathcal{U}}=1.9$, $\Lambda _{\mathcal{U}}=1000$
GeV, $C_{P}=C_{S}=2\times 10^{-3}$ and $C_{V}=C_{A}=10^{-5}$. Solid
line is for SM, dashed line is for scalar operator and long-dashed
line is for the vector operator.}\label{fig1}
\end{center}
\end{figure}


In case of $B\rightarrow K_{0}^{\ast }\left( K_{1}\right) \nu
\bar{\nu}$ there is no experimental limit on the branching ratio of
these decays, but these will be expected to be measured at future
SuperB factories where they analyze the spectra of final state
hadron by imposing a cut of on the high momentum of hadron to reduce
the background. To calculate the numerical value of the branching
ratio for $B\rightarrow K_{0}^{\ast }\left(
K_{1}\right) \nu \bar{\nu}$ in SM we have to integrate Eqs. (\ref{09}) and (%
\ref{12}) on the energy of the final state hadron. Thus after
integration, the values of the branching ratios in SM\ are:
\begin{eqnarray}
\mathcal{B}r\left( B\rightarrow K_{0}^{\ast }\nu \bar{\nu}\right)
&=&1.12\times 10^{-6}  \label{26} \\
\mathcal{B}r\left( B\rightarrow K_{1}\nu \bar{\nu}\right)
&=&1.77\times 10^{-6}  \notag
\end{eqnarray}%
With these values at hand, we have plotted the differential decay
with for $B\rightarrow K_{0}^{\ast }\left( K_{1}\right)
+\not\!\!{E}$ as a function of the energy of the final state hadron
$E_{K_{0}^{\ast }}\left(
E_{K_{1}}\right) $ and by fixing the parameters of unparticle from ref. \cite%
{22} in Fig.\ref{fig1}. One can easily see from the figure that the
signature of unparticle operators are very distinctive from the SM
for the final state hadron's energy. Just like $B\rightarrow K\left(
K^{\ast }\right) +\not\!\!{E}$ the distribution of unparticle
contribution is quite different when we include a vector operator
$\left( a=1\right) $ for the highly energetic final state hadron.
For the other values of $a$ we will discuss this issue separately.
Thus the Super B-factories will be able to clearly distinguish the
presence of unparticle by observing the spectrum of final state
hadrons in $B\rightarrow K_{0}^{\ast }\left( K_{1}\right)
+\not\!\!{E}$ in complement to $B\rightarrow K\left( K^{\ast
}\right) +\not\!\!{E}$.

\begin{figure}[h]
\begin{center}
\includegraphics[scale=0.7]{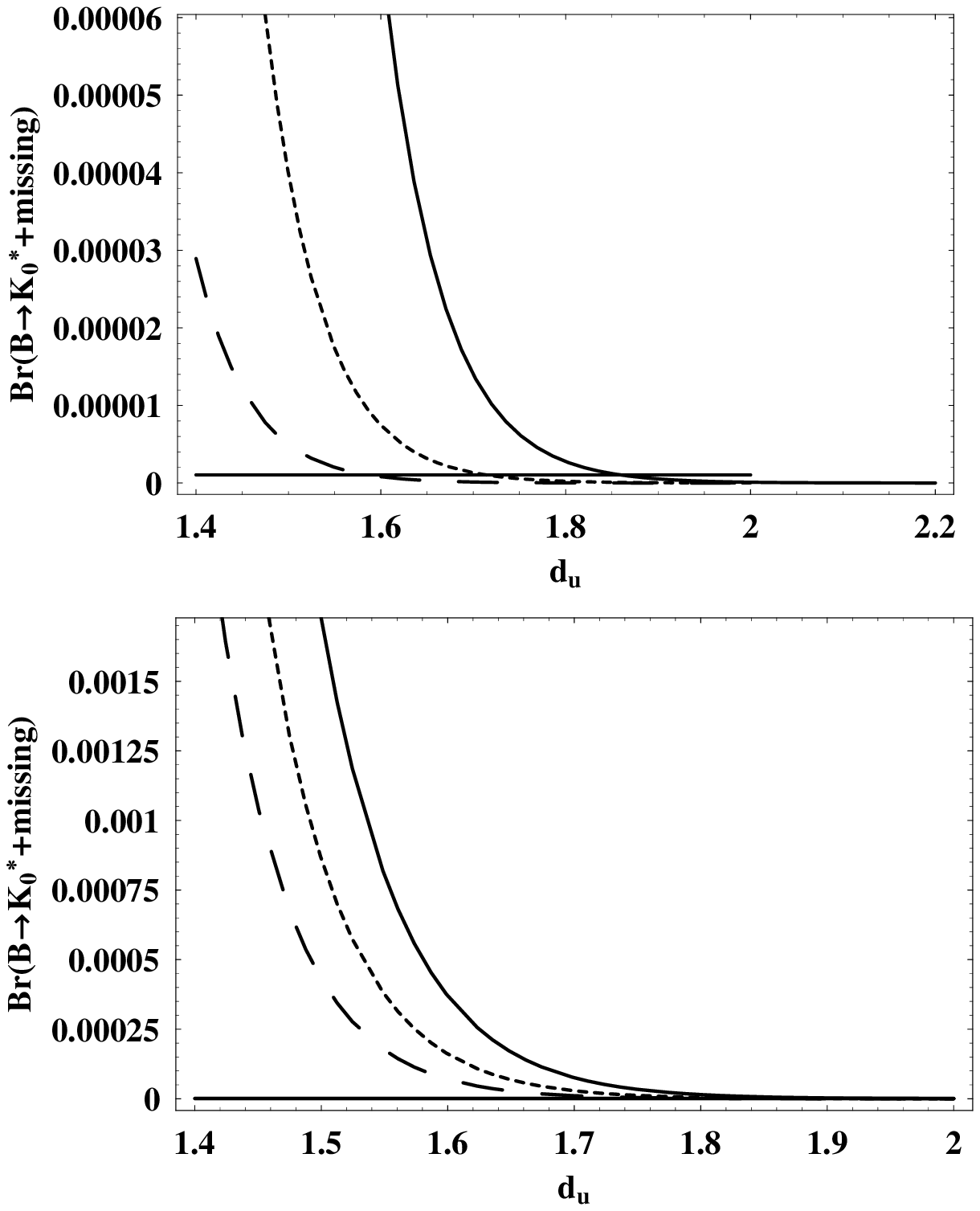}
\caption{The branching ratio for $B\rightarrow K_{0}^{\ast }+\not\!\!%
{E}$ as a function of $d_{\mathcal{U}}$ for various values of $\Lambda _{%
\mathcal{U}}$. Top panel is for the scalar operator and bottom is
for the vector operator. The values of coupling constants is same as
taken for Fig.
1. Solid line is for $\Lambda _{\mathcal{U}}=1000$ GeV, dashed line is for $%
\Lambda _{\mathcal{U}}=2000$ GeV and long-dashed line is for $\Lambda _{%
\mathcal{U}}=5000$ GeV$.$ The horizontal solid line is the SM\
result.}\label{fig2}
\end{center}
\end{figure}


\begin{figure}[h]
\begin{center}
\includegraphics[scale=0.7]{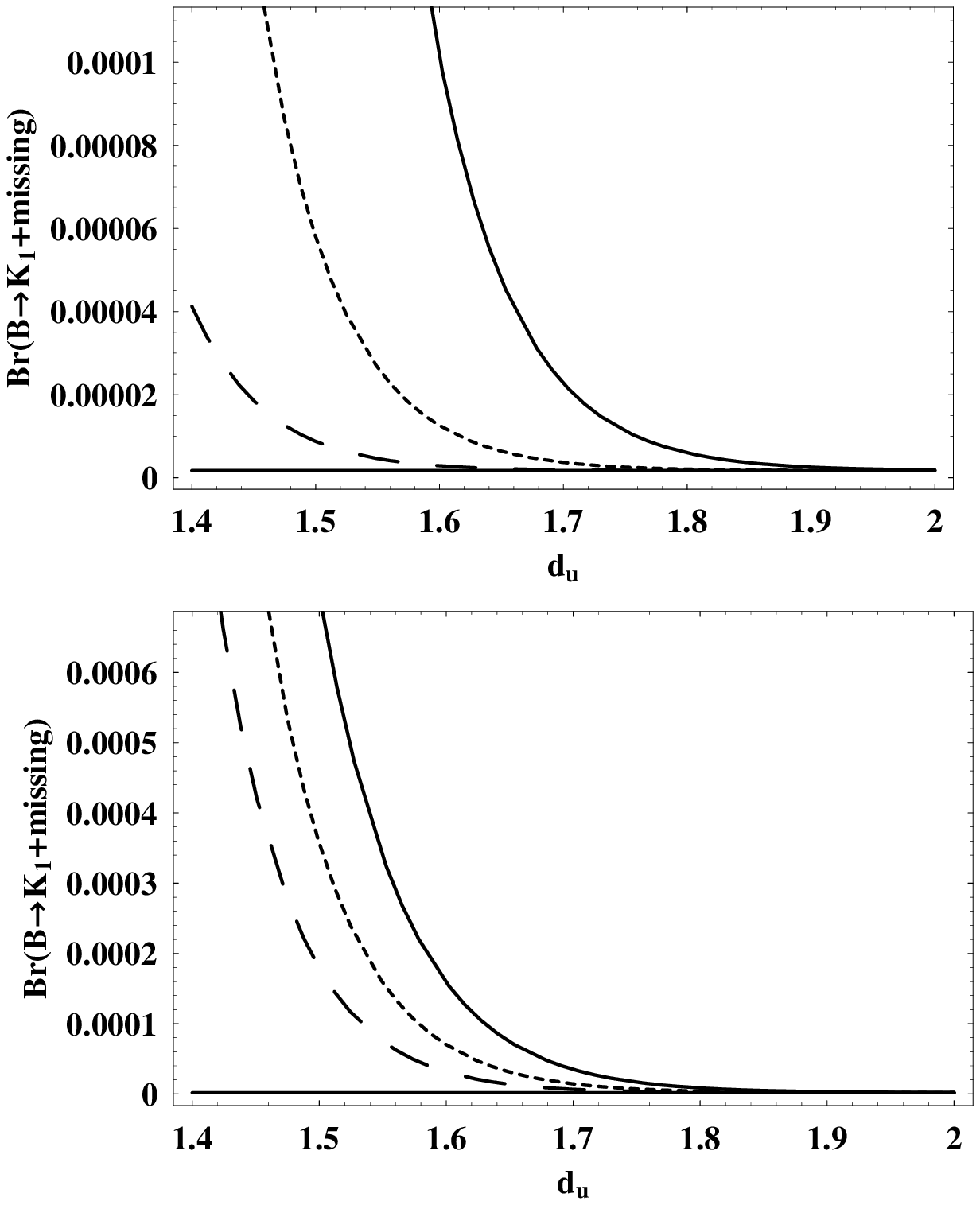}
\caption{The branching ratio for $%
B\rightarrow K_{1}+\not\!\!{E}$ as a function of $d_{\mathcal{U}}$
for various values of $\Lambda _{\mathcal{U}}$. Top panel is for the
scalar operator and bottom is for the vector operator. The values
for coupling constants is same as taken for Fig. 1. Solid line is
for $\Lambda _{\mathcal{U}}=1000$ GeV, dashed line is for $\Lambda
_{\mathcal{U}}=2000$ GeV and long-dashed line is for $\Lambda
_{\mathcal{U}}=5000$ GeV$.$ The horizontal solid line is the SM\
result.}\label{fig3}
\end{center}
\end{figure}

In Fig. 2 and Fig. 3 we have shown the sensitivity of the branching
ratio on the scaling dimension $d_{\mathcal{U}}$ for different
values of the cut-off scale $\Lambda _{\mathcal{U}}$ by using the
same values of $C_{S}$, $C_{P}$, $C_{V}$ and $C_{A}$ as we have used
for Fig. 1.\ We can see from this figure that the branching ratio is
very sensitive to the variable $d_{\mathcal{U}}$ and $\Lambda
_{\mathcal{U}}$. The constraints on the vector operator are more
stronger then the scalar operators and constraints for $B\rightarrow
K_{0}^{\ast }+\not\!\!{E}$ are better then the $B\rightarrow
K_{1}+\not\!\!{E}$ decays.

\begin{figure}[h]
\begin{center}
\includegraphics[scale=0.7]{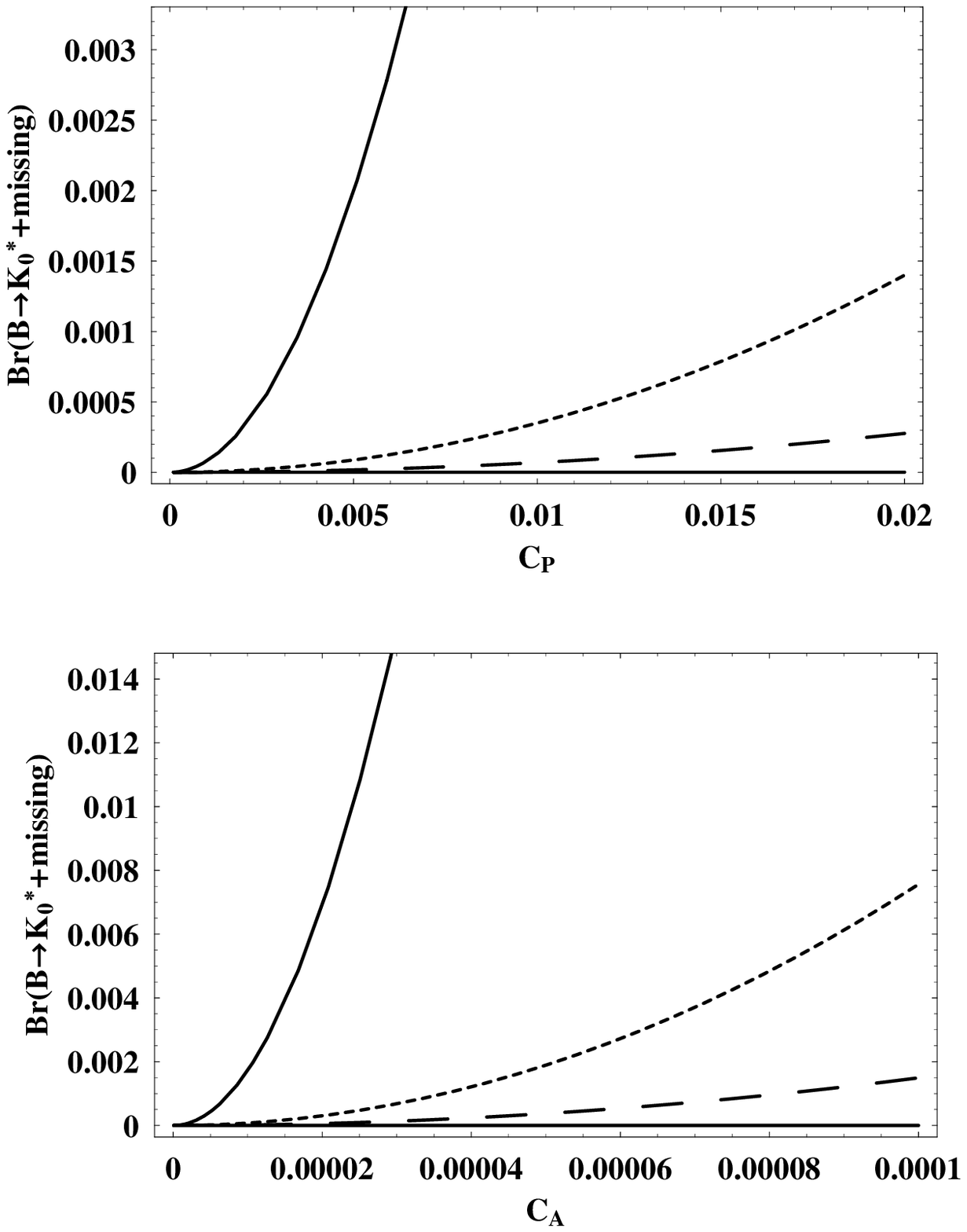}
\caption{The branching ratio for $%
B\rightarrow K_{0}^{\ast }+\not\!\!{E}$ as a function of $C_{P}$
(top panel) and
$C_{A}$ (bottom panel). The cut off scale has been taken to be $\Lambda _{%
\mathcal{U}}=1000$ GeV$.$ Solid line is for $d_{\mathcal{U}}=1.5$,
dashed
line is for $d_{\mathcal{U}}=1.7$ and long-dashed line is for $d_{\mathcal{U}%
}=1.9$. The horizontal solid line is the SM\ result.}\label{fig4}
\end{center}
\end{figure}

After showing the dependence of branching ratio on $d_{\mathcal{U}}$ and $%
\Lambda _{\mathcal{U}}$ what we have shown in Fig. 4 is the
sensitivity of the branching ratio of $B\rightarrow K_{0}^{\ast
}+\not\!\!{E}$ with the effective coupling constants of scalar and
vector unparticle operators. One can see that $B\rightarrow
K_{0}^{\ast }+$ scalar unparticle operator shall constrain the
parameter $C_{P}$ and $B\rightarrow K_{0}^{\ast }+$ vector
unparticle operator shall constrain the parameter $C_{A}$. Thus
observing this decay we can get some useful constraint on $C_{P}$
and $C_{A}$ which provides us the signature about the unparticle
physics. Similarly, we
have shown the dependence of the branching ratio of $B\rightarrow K_{1}+\not\!\!%
{E}$ on the effective coupling constants in Fig. 5. It is shown that
if we consider the scalar operator then only dependence is on
$C_{S}$, whereas if we consider the vector operators then the decay
rate depends both on $C_{V}$ and $C_{A}$.

\begin{figure}[h]
\begin{center}
\includegraphics[scale=0.7]{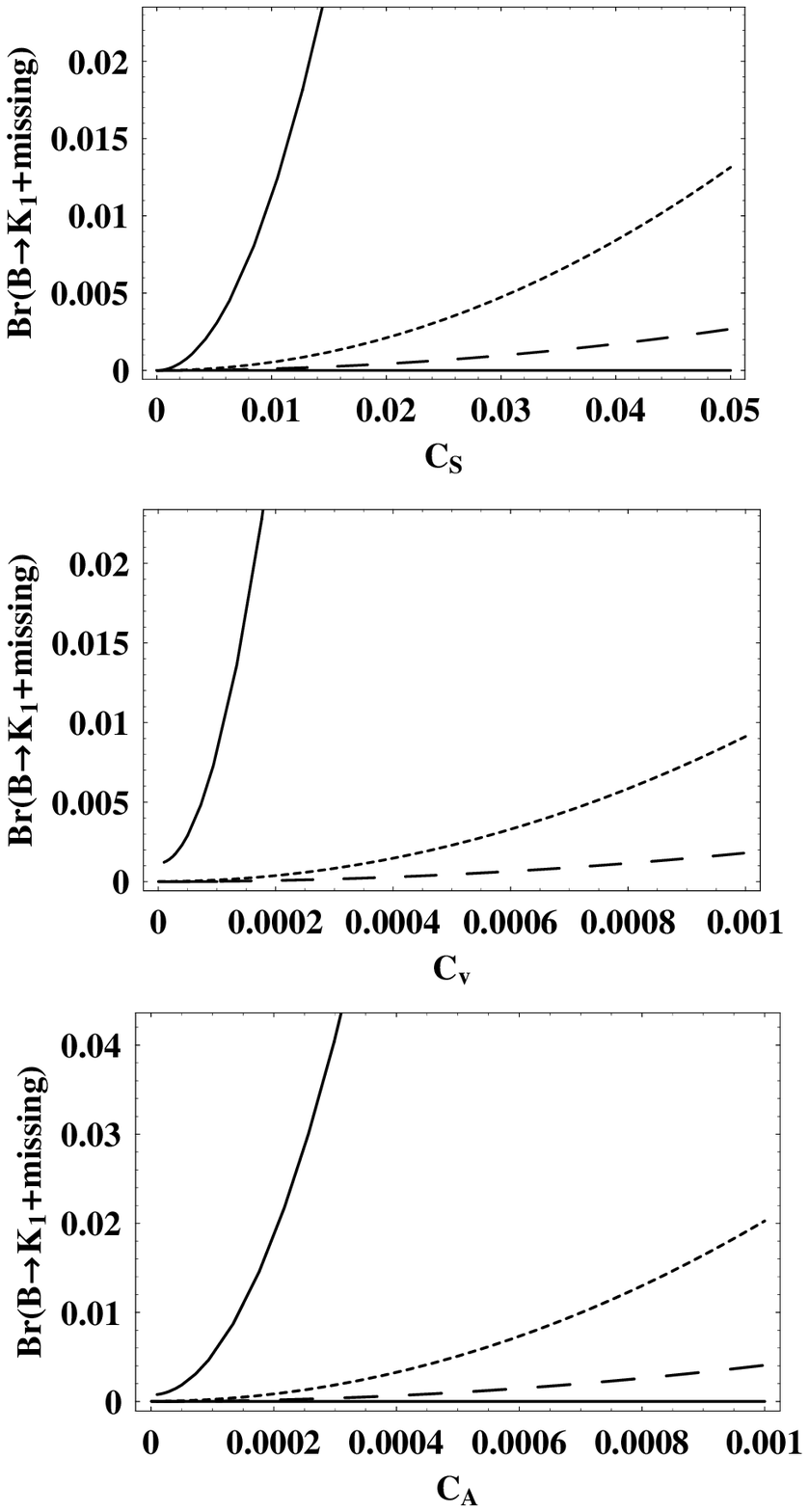}
\caption{The branching ratio for $B\rightarrow K_{1}+\not\!\!{E}$ as
a function of $C_{S}$ (top panel), $C_{A} $ (middle panel) and
$C_{V}$ (bottom panel). The cut off scale has been
taken to be $\Lambda _{\mathcal{U}}=1000$ GeV$.$ Solid line is for $d_{%
\mathcal{U}}=1.5$, dashed line is for $d_{\mathcal{U}}=1.7$ and
long-dashed line is for $d_{\mathcal{U}}=1.9$. The horizontal solid
line is the SM\ result.}\label{fig5}
\end{center}
\end{figure}

As we have already mentioned that, Grinstein et al. have recently given
their comment on the unparticle \cite{23} mentioning that one regards Mack's
unitarity constraint lower bounds on CFT operator dimensions, e.g., $d_{%
\mathcal{U}}\geq 3$ for primary, gauge invariant, vector unparticle
operators. To account for this they have corrected the results in the
literature, and modified the propagator of vector and tensor unparticles.
The modified expressions of decay rate for the processes under consideration
are given in Eq. (\ref{23}) and Eq. (\ref{24}). To incorporate this
modification in vector unparticle operator, what we have shown in Fig. 6 is
the fractional error

\begin{figure}[h]
\begin{center}
\includegraphics[scale=0.7]{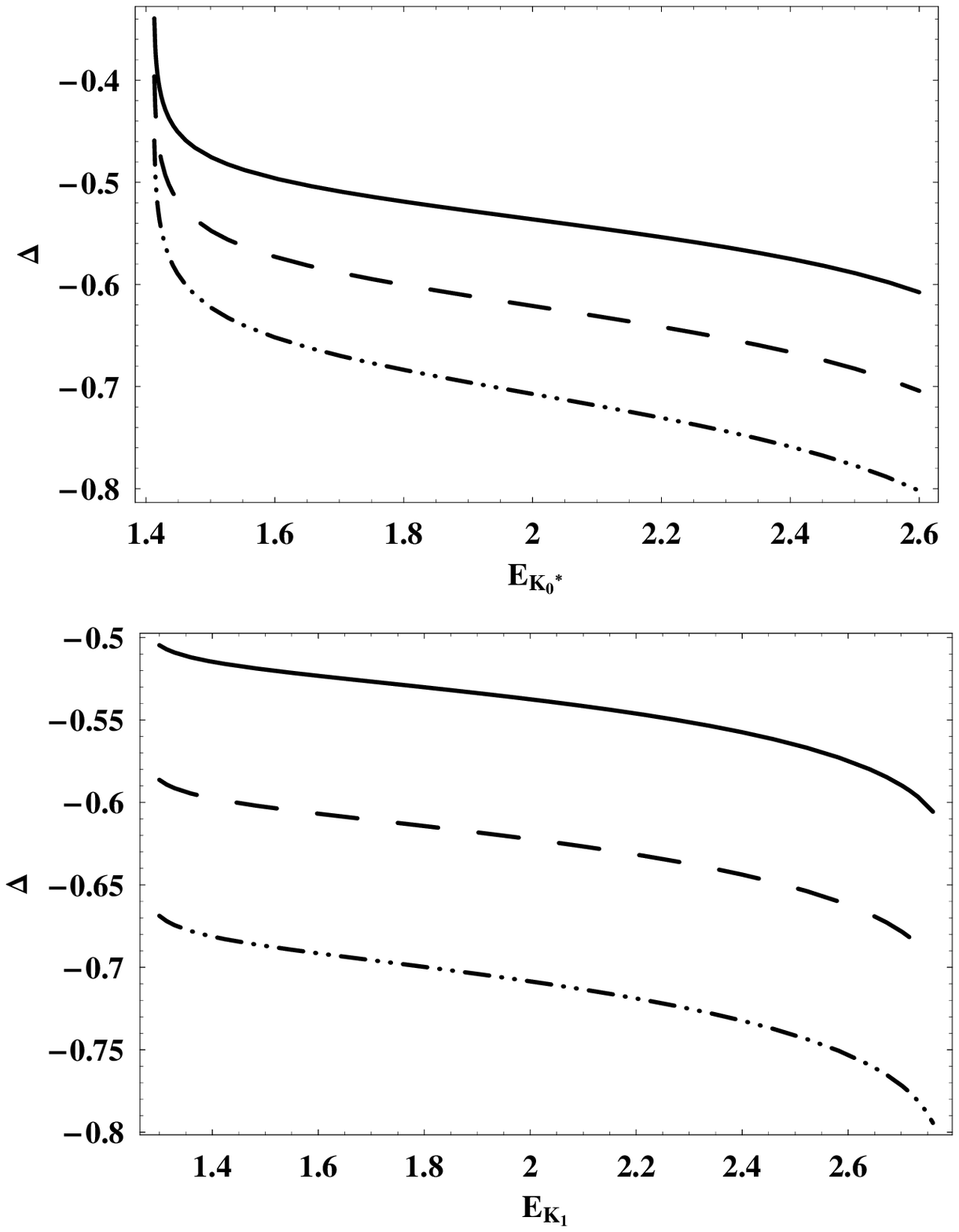}
\caption{Fractional error $\Delta $ in the spectrum for the decay
$B\rightarrow K_{0}^{\ast }\left( K_{1}\right) +$ vector unparticle
operator as a function of energy of final state hadron. Top panel
shows the value for $B\rightarrow K_{0}^{\ast }$ and the bottom
panel is for $B\rightarrow K_{1}$. The values for the coupling
constants and cutoff scale is same as taken for Fig. 1.
Solid line is for $d_{\mathcal{U}}=3.2$, dashed line is for $d_{\mathcal{U}%
}=3.4$ and dashed-double dotted is for
$d_{\mathcal{U}}=3.6$.}\label{fig6}
\end{center}
\end{figure}

\begin{equation}
\Delta \equiv \frac{\left( \frac{1}{\Gamma }\frac{d\Gamma }{dE_{K_{0}^{\ast
}\left( K_{1}\right) }}\right) _{a=1}-\left( \frac{1}{\Gamma }\frac{d\Gamma
}{dE_{K_{0}^{\ast }\left( K_{1}\right) }}\right) _{a}}{\left( \frac{1}{%
\Gamma }\frac{d\Gamma }{dE_{K_{0}^{\ast }\left( K_{1}\right) }}\right) _{a=1}%
}  \label{27}
\end{equation}%
where the difference is between the spectrum of $B\rightarrow K_{0}^{\ast
}\left( K_{1}\right) \mathcal{U}$ using vector unparticle operator with $a=1$
and $a=2\left( d_{\mathcal{U}}-2\right) /\left( d_{\mathcal{U}}-1\right) $
with $3<d_{\mathcal{U}}<3.9$. It is clear from the graph that if we increase
the unparticle scaling dimensions $d_{\mathcal{U}}$ the contribution of
vector unparticle operator to the decay rate decreases significantly because
of the increase in the inverse powers of cutoff scale $\Lambda _{\mathcal{U}%
} $ (see Eqs. (\ref{23}) and (\ref{24})).

Just to conclude: The study of these p-wave decays of $B$ mesons will not
only provide us information about SM but it also indicate the physics beyond
it and in future, when enough data is expected from the Super B factories,
we believe that these decays will take us step forward to the study of
unparticle as a source of missing energy in flavor physics.

\-\textbf{Acknowledgements:}

This work is partly supported by National Science Foundation of China under
the Grant Numbers 10735080 and 10625525. The authors would like to thank W.
Wang and Yu-Ming for useful discussions.

\end{document}